\documentclass[runningheads]{llncs}

\usepackage[T1]{fontenc}
\def\doi#1{\href{https://doi.org/\detokenize{#1}}{\url{https://doi.org/\detokenize{#1}}}}
\usepackage{graphicx}
\usepackage{color}
\usepackage{listings}
\begin{document}
\title{3D endoscopic depth estimation using \\3D surface-aware constraints}
\author{ Shang Zhao\inst{1} \and
  Ce Wang\inst{2,3} \and
  Qiyuan Wang\inst{1} \and
  Yanzhe Liu\inst{4} \and 
  S. Kevin Zhou\inst{1,2} }
 
\institute{
 Medical Imaging, Robotics, and Analytic Computing Laboratory and Engineering (MIRACLE) School of Biomedical Engineering \& Suzhou Institute for Advanced Research, University of Science and Technology of China, Suzhou, China  \and
 Key Lab of Intelligent Information Processing of Chinese Academy of Sciences (CAS), Institute of Computing Technology, CAS, Beijing, China \and
 Suzhou Institute of Intelligent Computing Technology, \linebreak Chinese Academy of Sciences, Suzhou, China  \and
 Faculty of Hepato-Biliary-Pancreatic Surgery, \linebreak Chinese People's Liberation Army (PLA) General Hospital, \linebreak Institute of Hepatobiliary Surgery of Chinese PLA, Beijing, China }

%
%
%
\maketitle              
\begin{abstract}
Robotic-assisted surgery allows surgeons to conduct precise surgical operations with stereo vision and flexible motor control. However, the lack of 3D spatial perception limits situational awareness during procedures and hinders mastering surgical skills in the narrow abdominal space. Depth estimation, as a representative perception task, is typically defined as an image reconstruction problem. In this work, we show that depth estimation can be reformed \textbf{from a 3D surface perspective}. We propose a loss function for depth estimation that integrates the \textbf{surface-aware constraints}, leading to a faster and better convergence with the valid information from spatial information. In addition, camera parameters are incorporated into the training pipeline to increase the control and transparency of the depth estimation. We also integrate a specularity removal module to recover more buried image information. Quantitative experimental results on endoscopic datasets and user studies with medical professionals demonstrate the effectiveness of our method.

\end{abstract}
%
%
%



\section{Introduction}

Endoscopic robotic-assisted surgery is a key breakthrough for the next-generation surgery room, which requires extensive research to realize the full intelligence of this emerging platform. Spatial perception, as a significant research topic for intelligent surgery, has the potential to empower spatial situational awareness with surgeons or surgical robots, resulting in better surgical outcomes~\cite{kassahun2016surgical,yang2017medical,long2022integrating,zhou2019handbook,zhou2021review}. Although the stereo endoscope camera views allow surgeons to estimate the depth with their eyes, it is difficult to avoid surgical action mistakes without objective spatial measurement. Therefore, standardizing spatial estimation with automation becomes a promising solution. In addition, depth estimation consolidates the intelligent surgery to provide spatial awareness with surgeons and support other useful applications such as surgical scene reconstruction~\cite{long2021dssr,recasens2021endo}.

However, calculating depth for endoscopic images faces various challenges. Endoscopic images contain complicated textural information, such as texture-less tissue and capillary pattern, hindering the model from learning effective feature descriptors. Researchers then formulate the depth estimation as an image reconstruction problem with pixels storing range values.
With the success of CNNs~\cite{liu2018self,guo2019group,laga2020survey,huang2021self} and Transformer~\cite{li2021revisiting}, further efforts have been made for better exploiting discriminative information that is valuable to depth estimation.

Nevertheless, endoscopic images suffer from rich specular regions that bury valid textural features due to smooth organ surfaces under endoscope illumination conditions. Previous representative solutions, such as the robust principal component analysis~\cite{da2014automatic,li2019specular}, have both robustness and computational efficiency issues. Although neural network solutions~\cite{chang2018single,fu2021multi} have been discussed in the computer vision domain, little work for endoscopy surgery has collaborated with other perception applications. To our best knowledge, we firstly integrate a specularity removal network module in the pipeline to improve the quality and efficiency of depth estimation.

Although acceptable performance improvements can be achieved from proper architecture and image restoration methods, constructing an effective loss function with full utilization of geometric information in the depth image is still one of the ultimate objectives. For this purpose,
the loss function has been widely researched since it can explicitly guide the model learning from specific image properties during training. For example, L1 pixel similarity and Structural Similarity Index Metric (SSIM) have dominated the training with image-space constraints~\cite{recasens2021endo,li2021revisiting}, which compare the depth pixel values and structural information respectively.
However, none of these works exploits 3D geometric features. Recently, image gradient-based loss has been discussed on 3D indoor scene datasets~\cite{hu2019revisiting,kusupati2020normal,long2021adaptive}, but the gradient computation cannot accurately indicate spatial geometric patterns in 3D camera space because their gradient computation merely relies on pixel values in 2D image space. \textbf{To bridge the gap, we formulate a new loss in 3D space with surface constraints that are computed by using camera parameters to perform back-projection on depth pixels for accurate 3D geometry recovery.}

In this work, we propose a modular pipeline for depth estimation. Specifically, a specularity removal module firstly recovers the buried image information from specular highlights. Then the Oriented Point (OP) Loss is utilized to bring in extra information and facilitate the training process, which formulates the constraints based on both point positions and orientations lying on 3D surfaces. 

The specific contributions of our work are the following:
\begin{itemize}
\item We develop an Oriented Point loss function that explicitly encodes 3D surface properties to assist depth estimation from endoscopic images. The point distance and normal similarity components in the OP loss constrain the surface boundary and local surface smoothness properties, respectively.

\item We encode the camera model formulation into the training pipeline, enabling an explicit control of depth estimation with more transparency.

\item We apply a specularity removal module to recover the texture information from specular highlights, facilitating feature extraction for stereo matching.

\item We conduct quantitative experiments and user studies to comprehensively evaluate the proposed method, exploring the potential value for practical clinical applications.
\end{itemize}

\section{Methods}

\begin{figure*}[t]
	\centering 
	\includegraphics[width=1.0\linewidth]{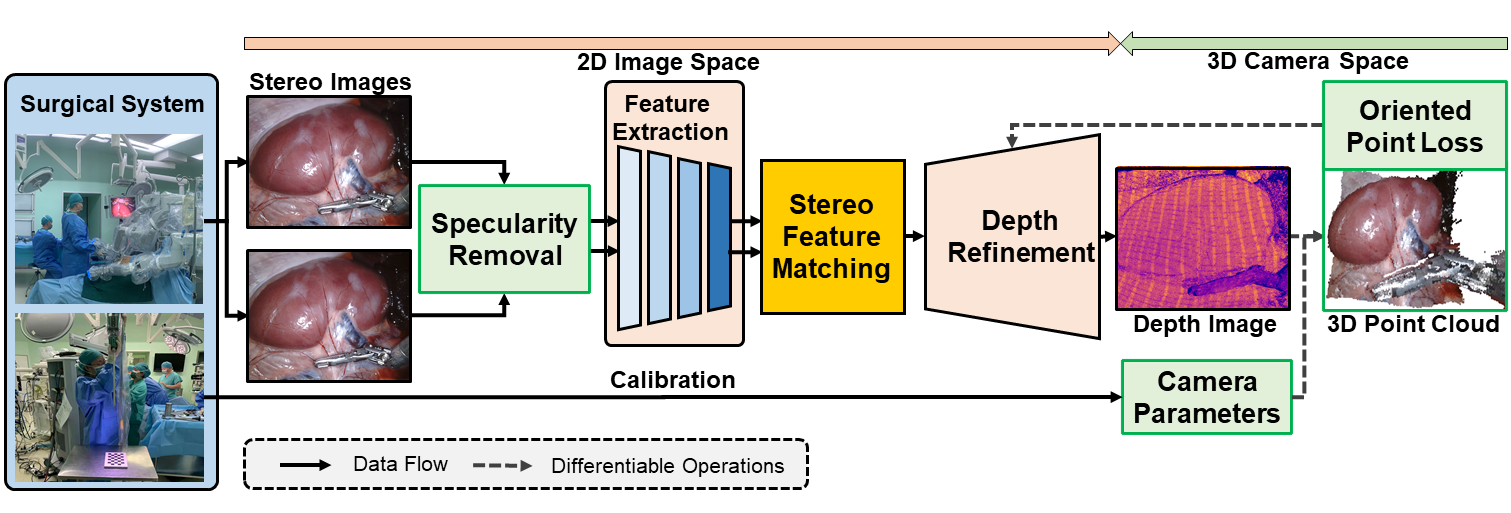}
	\caption{\textbf{The overview of the modular pipeline for depth estimation.} The entire pipeline reforms the traditional stereo matching pipeline with neural networks. The specularity removal module first recovers stereo images to retrieve more informative image content. Then a CNN feature extractor encodes the image with feature descriptors for stereo feature matching. The predicted depth is supervised by the corresponding oriented point samples produced by our proposed OP loss, where we retrieve 3D oriented samples with camera parameters. }
	\label{fig:overview}
\end{figure*}

Given a pair of stereo endoscopic images and the corresponding camera parameters $\mathbf{K} \in  {\rm I\!R}^{3\times3} $, we use a modular pipeline, as shown in Fig. \ref{fig:overview}, that can be implemented in an encoder-decoder fashion to predict the depth image $\hat{\mathbf{D}}$. 
In particular, endoscopic images are first processed with a specularity removal module to reveal more valid content. The OP loss then guides the model to learn 3D surface information to compensate for the information missing in the 2D image space.

\subsection{Depth Estimation from 3D Surface Perspective}
\subsubsection{Oriented Point Samples for 3D Surface Estimation.}
A point cloud is a lightweight representation for storing discrete surface information. Estimating 3D surfaces from point samples is a regression problem that aims to fit the optimal surface through the given samples, which is a well-studied problem in computer graphics~\cite{kazhdan2005reconstruction,kazhdan2006poisson}. The two most necessary geometry properties in these point samples are positions $\vec{p}$ and corresponding normals $\vec{n}$, which provide 3D surface boundary information and describe the local surface smoothness information around each point sample in Euclidean space, respectively.

Inspired by the point-based surface estimation, we revisit the depth image and treat it as a special 3D representation that inherently stores discrete point samples in the image space. The common perspective for depth image estimation in endoscopic images is to minimize the depth error between predicted and ground truth depth images. We move a step further and reform depth estimation as a 3D surface minimization problem with oriented point samples that learns the optimal mapping between 3D surfaces of prediction and ground truth.

\subsubsection{Point and Normal Computation from Depth Image.} 
The point cloud can be evaluated by depth image back-projection with camera parameters $\mathbf{K}$ from 2D image space, representing 3D surfaces with discrete samples in the 3D camera space. The involved back-projection mapping  ${\mathbf{\Gamma}(d_{\mathbf{u}},\mathbf{K})}^{-1}$ is defined as
\begin{equation}
{\mathbf{\Gamma}(d_{\mathbf{u}},\mathbf{K})}^{-1} = d_{\mathbf{u}}{\mathbf{K}}^{-1} \left[ x, y, 1 \right]^\mathsf{T},
\end{equation}
where $d_{\mathbf{u}}$ denotes the depth value of image pixel $\mathbf{u} \in {\rm I\!R}^{2} = (x, y)^\mathsf{T}$ in the image domain. Normals can be approximated with 3D point samples of neighboring pixel coordinates, which describe the outward-pointing orientation. Once we have the point cloud image, we compute 3D surface normals using the spatial gradient of the point cloud image by applying a Sobel-like filter $\Delta$ to the image~\cite{newcombe2011kinectfusion}. 
The normal estimation operator is defined as 
\begin{equation}
\mathbf{\Phi}({d}_{\mathbf{u}},N({d}_{\mathbf{u}}),{\mathbf{K}}) = \Delta_{x}(\mathbf{\Gamma}(d_{\mathbf{u}},\mathbf{K})^{-1})\times \Delta_{y}(\mathbf{\Gamma}(d_{\mathbf{u}},\mathbf{K})^{-1}), 
\end{equation}
where $\times$ denotes the vector cross product. Indeed, we can treat normal computation as a gradient form for 3D geometry information. Camera parameters are integrated into surface constraints since both point and normal computations are differentiable. Therefore, the transparency and control of depth estimation are improved by introducing constraints from 3D geometry.

\subsection{Oriented Point (OP) Loss Function}
Inspired by 3D geometry principles, we derived a two-part OP loss function, $\mathcal{L}_{OP}$, to improve the training process, including a point distance loss $\mathcal{L}_{p2p}$ and a normal similarity loss $\mathcal{L}_{n2n}$. The two parts provide extra metrics to evaluate surface properties that can be retrieved from the depth image, guiding convergence and imposing both geometry boundary and surface smoothness constraints:
\begin{equation}
\label{l_geom}
\mathcal{L}_{OP} = \alpha_{1} \mathcal{L}_{p2p} + \alpha_{2} \mathcal{L}_{n2n}.
\end{equation}

\begin{figure*}[t]
	\centering 
	\includegraphics[width=0.702\linewidth]{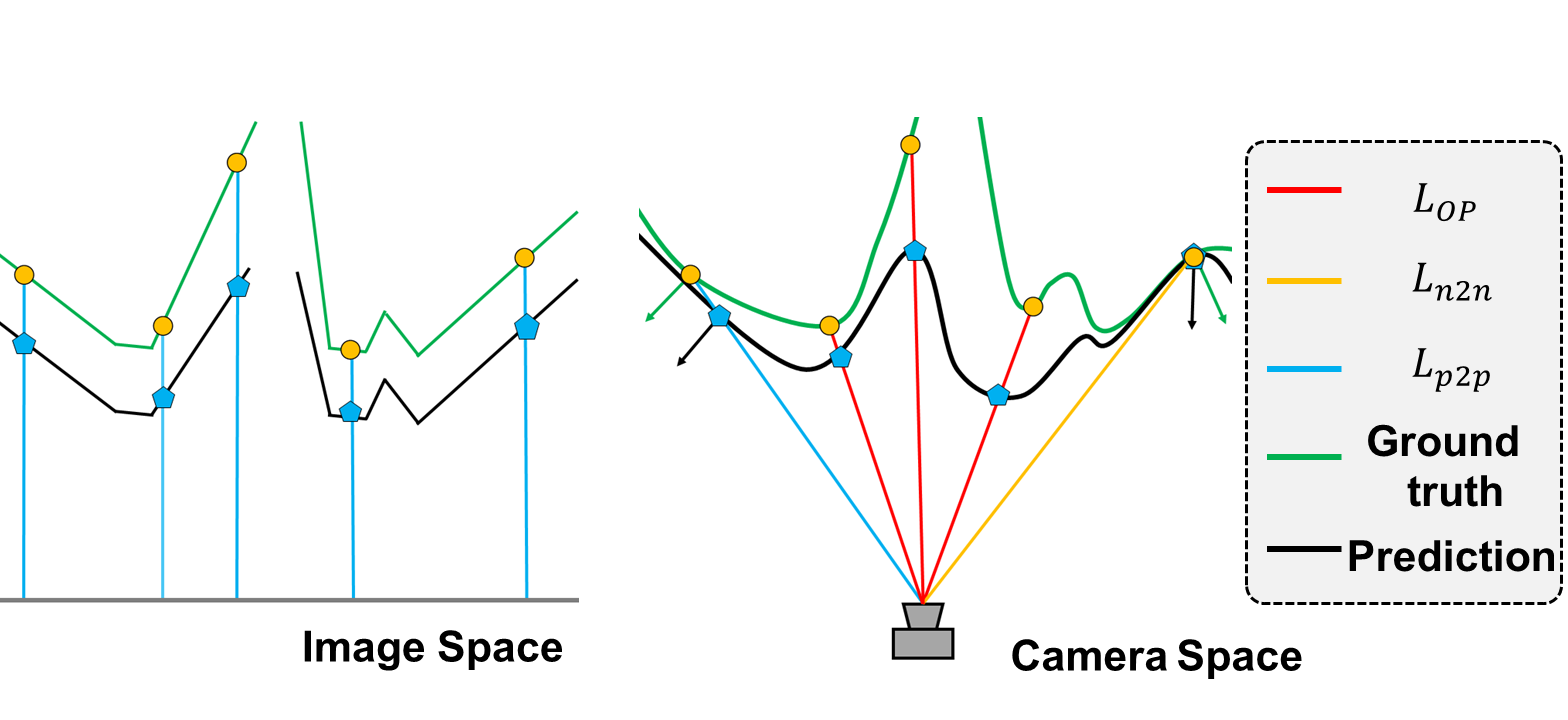}
	\caption{\textbf{A top-down view for image-based loss (Left) and 3D-based OP loss (right).} The pixel coordinates are aligned for the prediction (black) and ground truth (green). Image-based loss aggregates the pixel-to-pixel difference along orthogonal rays through pixels. In contrast, the OP loss jointly constrains the point distance and the spatial orientation that are perpendicular to the local surface.
	The blue line represents the example that has distance estimation in the corresponding space. The yellow line represents that the sample comparison only triggers orientation constraints. The red line shows the cases where both components contribute to the OP loss computation.}
	\label{fig:local_att_result}
\end{figure*}

In particular, the point distance loss aims to incorporate the spatial information into the optimization while preserving the depth value similarity metric on the z-dimensional component. In this way, each 3D point sample correlates with a depth pixel that shares the same image coordinate. This benefits from the spatial transformation controlled by camera parameters. The point distance loss is defined as below:
\begin{equation}\mathcal{L}_{p2p} = \sum_{\mathbf{u}}^{}\left\| \mathbf{\Gamma^{-1}}({\hat{d}_{\mathbf{u}}}, {\mathbf{K}} ) - \mathbf{\Gamma^{-1}}({d_{\mathbf{u}}}, {\mathbf{K}} ) \right\|_{1},
\end{equation}
where ${\mathbf{K}}$ denotes intrinsic parameters, $\mathbf{\Gamma^{-1}}$ denotes the back-projection operation.  ${\hat{d}_{\mathbf{u}}}$ and ${d_{\mathbf{u}}}$ denotes the predicted depth and the ground truth depth value at pixel $\mathbf{u}$, respectively.
To constrain the surface smoothness for depth estimation, we additionally apply a normal loss on endoscopic depth estimation that guides the model inference to preserve the similar surface smoothness of ground truth: 
\begin{equation}\mathcal{L}_{n2n} = \sum_{\mathbf{u}}^{}\left\| 1- \langle\mathbf{\Phi}({\hat{d}_{\mathbf{u}}}), \mathbf{\Phi}({d}_{\mathbf{u}})\rangle \right\|_{1} 
+ \beta\sum_{\mathbf{u}}^{}\left\| 1-\langle \mathbf{\Phi}({\hat{d}_{\mathbf{u}}}), \mathbf{\Phi}({\hat{d}_{\mathbf{u}}})\rangle\right\|_{1}, 
\end{equation}
where the normal estimation operator, $\mathbf{\Phi}({\hat{d}}_{\mathbf{u}}) = \mathbf{\Phi}({\hat{d}}_{\mathbf{u}},N(\hat{d}_{\mathbf{u}}),{\mathbf{K}})$, 
computes the normal with differentiation, and $N(\hat{d}_{i})$ represents the neighboring pixels for normal computation. In more detail, the first component in $\mathcal{L}_{n2n}$ encourages the orientation alignment between the prediction and the ground truth. The other component regularizes the normal by penalizing the samples with large local spatial differences, which can be individually controlled by a scalar weight $\beta$.
Note that we define the similarity of the dot product operator as concerning the length of the unit vector. This loss component also implicitly encourages the similarity to be measured in the unit space. Fig.~\ref{fig:local_att_result} illustrates the differences between image-based loss and our 3D-based OP loss.

\subsection{Specularity Removal}
To reduce specular reflection in images, we adopted the network architecture from Fu et al.~\cite{fu2021multi} as the image enhancement stage to recover the image content from specular regions. This step improves the perception ability of the feature extraction module, enhancing the depth quality in texture-less specular regions.


\section{Experiments}

\noindent\textbf{Dataset.} We use the SCARED dataset~\cite{allan2021stereo} and the Hamlyn dataset~\cite{recasens2021endo} to evaluate the effectiveness of our proposed approach. SCARED is a laparoscopic scene dataset, consisting of stereo images with the resolution of 1280x1024, which needs to be preprocessed to retrieve rectified images. We apply a bilateral filter to remove noisy pixels. The Hamlyn dataset is the other dataset used in our experiments, which contains multiple intracorporeal scenes that are recorded with rectified stereo images and corresponding depth images with the resolution of 640x480. For both datasets, we randomly select about 2K images as training data.
The testing data for SCARED includes the 3 keyframe sequences (2200 images) in the test dataset 1, and the one for the Hamlyn dataset is the test sequence in the pre-defined split file based on rectified sequences 6, 8, and 9 (1000 images).

\noindent\textbf{Training Details and Evaluation Metrics.} 
We implement our modular depth estimation pipeline with PyTorch~\cite{paszke2019pytorch} and test models on an NVIDIA GTX 3090 GPU.
The lightweight stereo-transformer (STTR)~\cite{li2021revisiting} is used as the back-bone depth estimation module in our pipeline. 
Note that the poor density of depth maps in those endoscopic datasets results in a challenging fact that may aggravate the optimization without proper initialization. Inspired by STTR~\cite{li2021revisiting}, we use the SceneFlow dataset~\cite{MIFDB16} to pretrain the model with a small number of epochs to avoid the aforementioned situation. Then we fine-tune the models with SCARED and Hamlyn datasets for 50 epochs. In these training stages, we empirically set the OP loss weight to 0.05 and have trained all models with the AdamW optimizer~\cite{loshchilov2017decoupled} (lr=1e-5) without parameter searching. 

For validating the model performance, we use mean average error (MAE) and root mean square error (RMSE) as quantitative evaluation metrics.
We also analyze the ranking scores collected from experienced surgeons to quantify the visual quality and verify the practical usage of our results.



\subsection{Ablation Study}



\begin{table}[t]
 \caption{The quantitative results of the depth estimation on the SCARED and Hamlyn datasets. The MAE and RMSE in mm are the smaller the better.}

 \label{tab:sttr_scared}
 \centering
 \begin{tabular}{l|cccc|cccc | c|c}
  \hline
  \hline 
  Dataset & \multicolumn{8}{c|}{SCARED} & \multicolumn{2}{c}{Hamlyn} \\
  \hline
  Metric &\multicolumn{4}{c|}{{MAE(mm)($\downarrow$)}} & \multicolumn{4}{c|}{{RMSE(mm)($\downarrow$)}} & MAE($\downarrow$) & RMSE($\downarrow$) \\
  \hline
  Split & KF1 & KF2 & KF3 & ALL & KF1 & KF2 & KF3 & ALL & Test1 & Test1 \\
  \hline
 STTR \cite{li2021revisiting} &  57.05 & 76.34 & 78.37 & 68.95 & 57.46 & 76.79 & 78.46 & 69.38  &  72.09 & 89.55  \\
 STTR+$\mathcal{L}_{p2p}$ &  22.31 & 43.85 & 45.56 & 35.43 & 26.57 & 46.19 & 47.74 & 38.52  &  39.60 & 68.83  \\
 STTR+$\mathcal{L}_{n2n}$ & 34.48 & \textbf{18.06} & \textbf{16.95} & 24.54 & 37.85 & \textbf{22.55} & \textbf{21.48} & 28.57  & \textbf{38.24} & 76.25  \\
 STTR+$\mathcal{L}_{OP}$ & \textbf{18.72} & 21.28 & 21.60 & \textbf{20.33} & \textbf{26.35} & 25.40 & 25.82 & \textbf{25.92} & 43.36 & \textbf{45.43}\\
  \hline
  \hline
 \end{tabular}
\end{table}

\begin{table}[h]
 \caption{The quantitative results of specularity removal on the SCARED dataset. Note that we selected a subset that only includes the images with strong specular regions for each sequence. \textbf{SPR} denotes the specularity removal module.}
 \label{tab:spr_scared}
 \centering
 \begin{tabular}{l|cccc|cccc}
  \hline
  \hline
  Experiment&\multicolumn{4}{c|}{{MAE(mm)($\downarrow$)}}&\multicolumn{4}{c}{{RMSE(mm)($\downarrow$)}} \\
  \hline
   & KF1 &KF2&KF3 &ALL& KF1 &KF2&KF3 &ALL \\
  \hline
 STTR+$\mathcal{L}_{OP}$ \textbf{(w/o SPR)} & 22.08 &	42.65 &	45.62 &	35.01 &	24.66 &	44.45 &	47.38 &	37.13 \\
 STTR+$\mathcal{L}_{OP}$ \textbf{(w/ SPR)} & 20.94 & 37.51 & 39.46 & 31.22 & 27.47 &	41.39 &	42.94 &	36.08 \\
  \hline
  \hline
 \end{tabular}
\end{table}

We first conduct an ablation study with STTR on SCARED and Hamlyn datasets to evaluate the effectiveness of the proposed OP loss. Experimental results on SCARED dataset are shown in Table \ref{tab:sttr_scared}, including STTR as baseline, STTR with point distance loss, STTR with normal similarity loss, and the proposed OP loss. The results show that each part of the OP loss can individually improve the performance and that both components can jointly contribute to the performance as well. Although STTR with normal similarity achieves the best performance on two test keyframe sequences, STTR with OP loss achieves the best performance on the full test set, which outperforms by 48.62mm in MAE and 43.46mm in RMSE. We also observe a similar trend on Hamlyn dataset in Table \ref{tab:sttr_scared} that all parts of OP loss function positively impact the model performance. Both point distance and normal similarity outperform the OP loss on MAE, but they have less satisfactory results on RMSE. Note that the proposed OP loss outperforms by 44.12mm in RMSE, which indicates that the OP loss is effective in achieving a stable depth estimation with respect to noisy samples. All the results show that our loss can effectively improve the performance with 3D surface constraints.


To validate the specularity removal, we select the images with significant specular effects to construct a subset for each keyframe sequence in the SCARD test dataset. The results in Table \ref{tab:spr_scared} demonstrate that the model with specularity removal outperforms by 6.16mm in MAE and 4.44mm in RMSE on keyframe 3 and by 3.79mm in MAE and 1.05mm in RMSE on the full test set, indicating that the integration of the specularity removal module indeed recovers the valid image content for the undesired specular regions. Encouraged by the performances with our plug-in pre-trained specularity-removal module, we will explore finetuning it simultaneously with other parts in the future, which would further help compensate for the difference between the intermediate goal of enhancing stereo images and the final objective of predicting depth images.

\subsection{Qualitative Analysis and User Study}
\begin{figure*}[t]
	\centering 
	\includegraphics[width=0.95\linewidth]{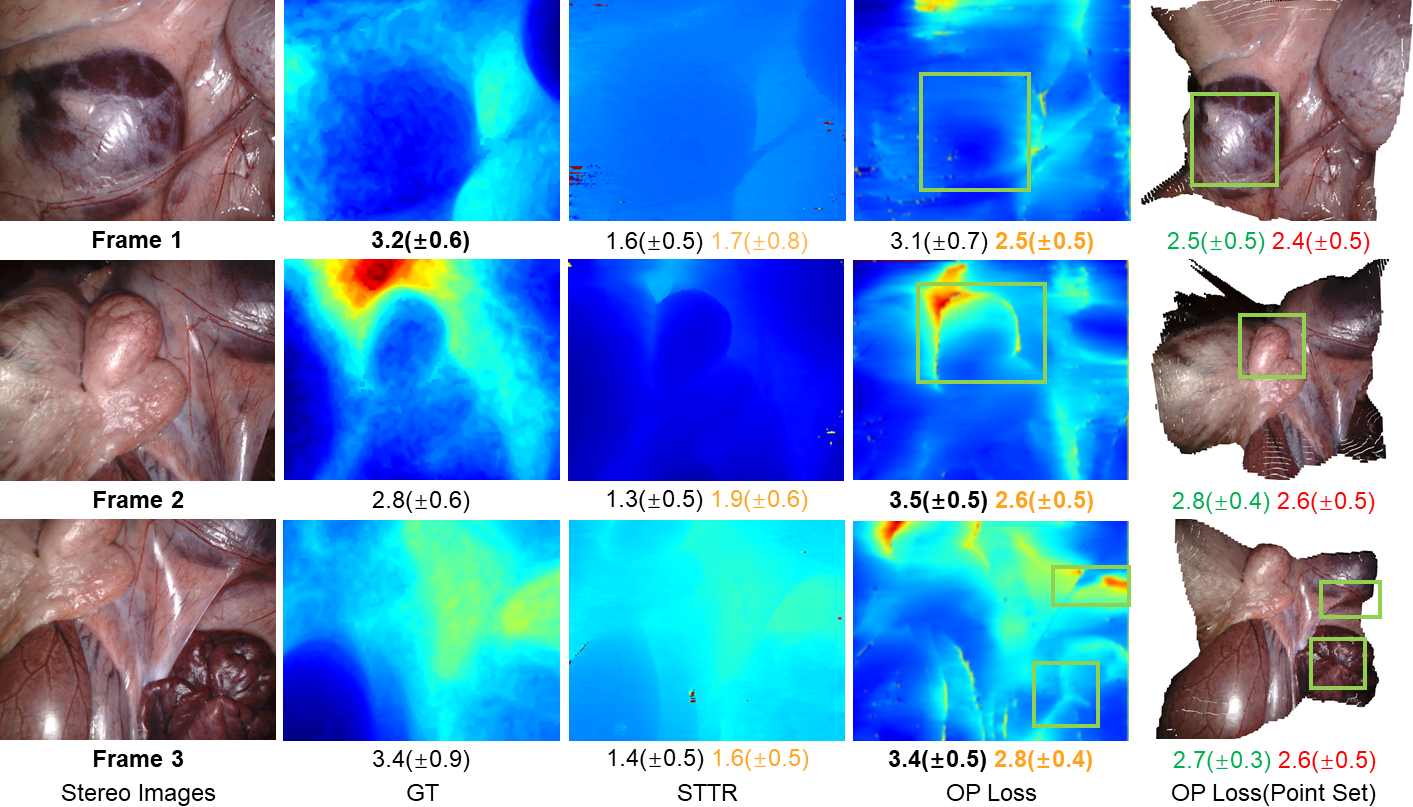}
	\caption{The qualitative results from the SCARED dataset. The warmer color represents a larger depth value. The OP loss guides the model to detect more detailed surface information, especially at the curves near the boundary. Our result is closer to the ground truth. The gradual change of color in OP loss prediction indicates better curvature detection on local surface patches. Both mean and standard deviation are provided for user study. The higher score indicates better results.}
    \label{fig:user-study}
\end{figure*}

To demonstrate the effectiveness of the proposed method, we further show qualitative comparisons of three representative prediction examples from SCARED dataset in Figure \ref{fig:user-study}. The results show that OP loss is capable of recovering more details in the depth perception, especially in the regions bounded by the green boxes. For example, the bottom right box in the OP loss prediction for frame 3 shows that our model recovers some fine-grained detail that is not even observed in the ground truth image. Although too many details may reflect a few artifacts, such as frame 1, our predictions are visually pleasurable in 3D. 

In addition, we conduct a user study to validate our method by collecting the questionnaires of 10 subjects who are more than 5-year experienced surgeons from our collaborated healthcare institution. The questionnaire exhibits several qualitative results, including the three ones in Figure \ref{fig:user-study}, along with 3D videos for each sample to demonstrate the depth prediction with spatial visualization. We first hide all image information to ask subjects to score correlation levels between RGB and depth images with a 4-point scale. The results are reported as black numbers, which show that our prediction has a close or better spatial description than ground truth images because it contains more detailed color gradients. Note that our prediction in frame 2 has a score that is higher than the ground truth by 0.7. Other colored numbers are on a 3-point scale. Orange numbers denote the similarity between predictions and the ground truth, which shows that our prediction has a better agreement on the similarity with the ground truth. Green and red numbers denote the agreements on spatial perception and colored point cloud quality, respectively, both confirming the effectiveness of our method.

\section{Conclusion}
In this paper, we have presented a modular depth estimation pipeline for endoscopic images, especially an OP loss that introduces 3D surface properties to facilitate depth estimation. Our OP loss jointly combines the surface boundary and smoothness to constrain the depth estimation as surface registration with oriented points. The specularity removal module recovers valid information from degraded images. Experimental results substantiate that our algorithm achieves better performance via utilizing 3D geometric information and restoring images. Future works include end-to-end integration of specularity removal into the framework and exploring the use of more 3D surface properties, such as isocontours, for better 3D endoscopic depth estimation.

\bibliographystyle{splncs04}
\bibliography{main}
%




\end{document}